\documentclass[showpacs,amsmath,amssymb,aps,showkeys,floatfix,prd,a4paper]{revtex4}

\usepackage[dvips]{graphicx}
\usepackage{dcolumn}
\usepackage{bm}
\usepackage{epsfig}
\usepackage{amsfonts}
\usepackage{amssymb,amscd}

\def\lsim{\raise0.3ex\hbox{$<$\kern-0.75em\raise-1.1ex\hbox{$\sim$}}}

\def\gsim{\raise0.3ex\hbox{$>$\kern-0.75em\raise-1.1ex\hbox{$\sim$}}}

\def\pom{{I\!\!P}}

\newcommand{\be}{\begin{equation}}

\newcommand{\ee}{\end{equation}}

\def\beq{\begin{equation}}

\def\eeq{\end{equation}}

\def\beqa{\begin{eqnarray}}

\def\eeqa{\end{eqnarray}}

\newcommand{\rd}{\mbox{\boldmath $\Delta$}}

\newcommand{\ba}{\begin{eqnarray}}

\newcommand{\rr}{\mbox{\boldmath $r$}}

\newcommand{\rb}{\mbox{\boldmath $b$}}

\def\gappeq{\mathrel{\rlap {\raise.5ex\hbox{$>$}}

{\lower.5ex\hbox{$\sim$}}}}

\def\lappeq{\mathrel{\rlap{\raise.5ex\hbox{$<$}}

{\lower.5ex\hbox{$\sim$}}}}

\def\Toprel#1\over#2{\mathrel{\mathop{#2}\limits^{#1}}}

\def\pom{{I\!\!P}}

\begin{document}

\begin{flushright}
MS-TP-23-19
\end{flushright}

\title{Double particle production in ultraperipheral $PbPb$ collisions \\ at the Large Hadron Collider and Future Circular Collider}

\author{Celsina N. {\sc Azevedo}}
\email{acelsina@gmail.com}
\affiliation{Physics and Mathematics Institute, Federal University of Pelotas, \\
  Postal Code 354,  96010-900, Pelotas, RS, Brazil}

\author{Victor P. {\sc Gon\c{c}alves}}
\email{barros@ufpel.edu.br}
\affiliation{Institut f\"ur Theoretische Physik, Westf\"alische Wilhelms-Universit\"at M\"unster,
Wilhelm-Klemm-Straße 9, D-48149 M\"unster, Germany
}
%\affiliation{Institute of Modern Physics, Chinese Academy of Sciences,
%  Lanzhou 730000, China}
\affiliation{Physics and Mathematics Institute, Federal University of Pelotas, \\
  Postal Code 354,  96010-900, Pelotas, RS, Brazil}

\author{Bruno D. {\sc Moreira}}
\email{bduartesm@gmail.com}
\affiliation{Departamento de F\'isica, Universidade do Estado de Santa Catarina, 89219-710 Joinville, SC, Brazil.}

\begin{abstract}
In this paper we analyze the associated production of a vector meson with a (pseudo)scalar bound state or a dimuon system in ultraperipheral $PbPb$ collisions through the double scattering mechanism for the energies of the Large Hadron Collider (LHC) and Future Circular Collider (FCC). Our results complement  previous studies for the double vector meson production. We present our predictions for the total cross sections and rapidity distributions considering the rapidity ranges covered by the ALICE and LHCb detectors, which indicate that a future experimental analysis of the $\phi + \eta_c$,  $\phi + \mu^+ \mu^-$ and  $J/\Psi + \mu^+ \mu^-$ final states is feasible. 
\end{abstract}



\maketitle

\vspace{1cm}

%\section{Introduction}
Ultraperipheral heavy ion collisions (UPHICs) provide an opportunity to  improve our understanding of the Standard Model as well as to searching for New Physics \cite{upc}. Over the last decades, a large number of  theoretical studies have been performed, mainly focused on  exclusive processes, which can be described in terms of the interaction between color singlet objects, e.g.  photon ($\gamma$) and/or Pomeron ($\pom$), and that are characterized  by  the incident ions remaining intact and two rapidity gaps being generated in the final state. Typical examples of exclusive processes are the production of (pseudo)scalar mesons by photon -- photon ($\gamma \gamma$) interactions and vector mesons by photon -- Pomeron ($\gamma \pom$) interactions, which allow us investigate the description of the meson wave functions as well as the treatment of the QCD dynamics at high energies (See, e.g. Refs. \cite{klein,gluon,Frankfurt:2001db,Bertulani:2009qj,Goncalves:2012zm,Moreira:2016ciu,run2,Goncalves:2018hiw,Goncalves:2021ytq}). Moreover, the production of two vector mesons in UPHICs by $\gamma \gamma$ interactions has also be considered, mainly motivated by the perspective of use this process to probe the QCD Pomeron (See, e.g. Refs. \cite{Goncalves:2002vq,Goncalves:2005rz,Goncalves:2006hu,Klusek:2009yi,Baranov:2012vu,Goncalves:2015sfy}). More recently, the double meson production by $\gamma \pom$ interactions in $pA$ collisions      has been investigated in Refs. \cite{Andrade:2022wph,Andrade:2022rbn}. 
A basic assumption in these  analyzes is that the final state system is produced by a single $\gamma \gamma$ or $\gamma \pom$ interaction. However, as pointed out originally in 
Ref. \cite{klein}, the probability of two simultaneous exclusive interactions is non - negligible when one has the collision of two heavy ions at high energies due to the huge number of photons emitted by the incident ions ($\propto Z^2$).  Such a conclusion was confirmed by the detailed studies of the  double scattering mechanism (DSM) performed in Refs. \cite{Klusek-Gawenda:2013dka,DSM,Azevedo:2022ozu} for the photoproduction of two vector mesons through two  $\gamma \pom$ interactions   and in Refs.  \cite{Artemyev:2014eaa,Klusek-Gawenda:2016suk,vanHameren:2017krz,Karadag:2019gvc}    for the production of two dilepton pairs by two $\gamma \gamma$ interactions. In particular, the results presented in Refs. \cite{Klusek-Gawenda:2013dka,DSM,Azevedo:2022ozu}  indicate that the study of the  associated production of two vector mesons can be considered an alternative to improve our understanding of the QCD dynamics as well as can be used for testing the treatment of the double scattering mechanism in UPHICs. Fortunately, the experimental analysis of the $\phi J/\Psi$ production is  currently being considered by the LHCb Collaboration and the data releasing is expected in the forthcoming years \cite{private}. 

% the particle production of distinct final state systems that can be studied in UPHICs, as e.g. the production of dileptons, diphotons, vector mesons, exotic systems and new particles (axion-like particles, dark photons, ....). A particular emphasis has been given in the study of exclusive processes, which can be described in terms of the interaction between color single objects, e.g.  photon ($\gamma$) and/or Pomeron ($\pom$), and that are characterized  by  the incident ions remaining intact and two rapidity gaps being generated in the final state
%In Refs. \cite{Klusek-Gawenda:2013dka,DSM,Azevedo:2022ozu,Artemyev:2014eaa,Klusek-Gawenda:2016suk,vanHameren:2017krz,Karadag:2019gvc}, the authors have considered that the final state is generated by the double scattering mechanism through two $\gamma \pom$ {\it or} two $\gamma \gamma$ interactions that occur simultaneously. 

\begin{figure}[t]
\includegraphics[scale=0.4]{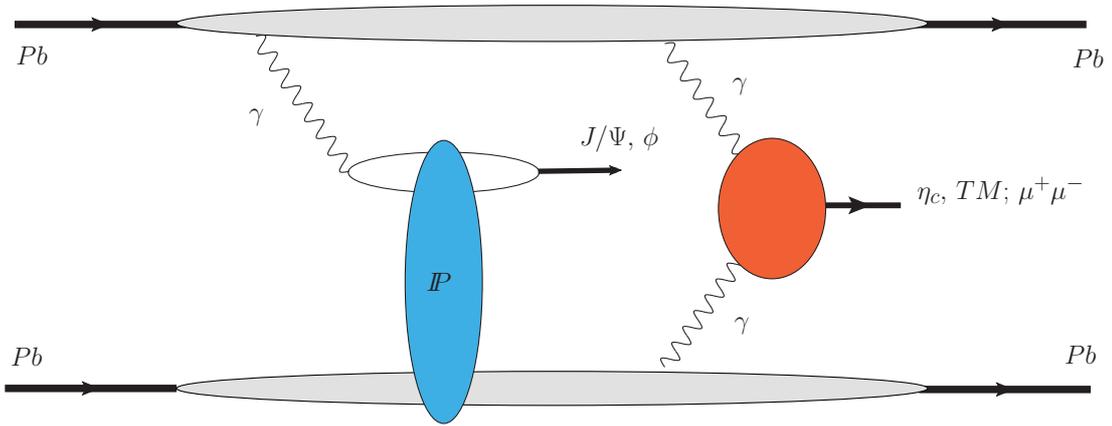}
\caption{Double particle production by $\gamma \pom$ {\it and} $\gamma \gamma$ interactions in ultraperipheral $Pb Pb$ collisions.}
\label{Fig:diagram}
\end{figure}

In this letter, we will investigate, for the first time, the double particle production in UPHICs when the final state is generated  by  $\gamma \pom$ {\it and}  $\gamma \gamma$ interactions, as represented in Fig. \ref{Fig:diagram}. One has that if these interactions occur, a vector meson ($V$) can be produced in association with a (pseudo)scalar state or  a dilepton pair ($l^+l^-$). In order to complement the results derived in Ref. \cite{Azevedo:2022ozu} we will assume in  our analysis that $V = J/\Psi$ or $\phi$. We will consider two different bound states that can be produced by $\gamma \gamma$ interactions: the $\eta_c$ meson and the true muonium ($TM$), which is the $\mu^+ \mu^-$ bound state with  total spin $s = 0$ (denoted para-$TM$ and represented by $(\mu^+ \mu^-)$ in Ref. \cite{Azevedo:2019hqp}).
The analysis considering a TM state is strongly motivated by the recent results presented in Refs. \cite{Azevedo:2019hqp,Francener:2021wzx,dEnterria:2022ysg,Yu:2022hdt,dEnterria:2023yao}, which have demonstrated that the study of QED bound states in UPHICs are  an ideal testing ground of QED and a sensitive probe of New Physics.
Moreover, we will consider that $l^+l^- = \mu^+\mu^-$, motivated by the fact we  are able to describe the current data for the  dimuon production in a single $\gamma \gamma$ interaction (See Ref. \cite{Azevedo:2019fyz}).
In our analysis, we will consider $PbPb$ collisions for the energies of the LHC ($\sqrt{s} = 5.02$ and 5.5 TeV) and FCC ($\sqrt{s} = 39$ TeV) and    estimate the 
 total cross sections and rapidity distributions for the double particle production assuming the rapidity ranges  covered by the ALICE ($-2.5 \le Y \le 2.5$) and LHCb ($2.0 \le Y \le 4.5$) detectors. As we will demonstrate below, a future experimental analysis of the associated production of a vector meson with $\eta_c$ meson or a dimuon system is, in principle, feasible.

Initially, let's present a brief review of the formalism needed to describe the double scattering mechanism (DSM) in UPHICs (For details see, e.g., Refs. \cite{klein,DSM}). The basic assumption in the current formalism to treat the double particle production via DSM is that possible correlations can be neglected in a first approximation. As a consequence, one has that the pair production probability can be expressed in terms of the product of single production probabilities, which implies that  the double differential cross section for the production of a particle ${\cal{X}}$  at rapidity $y_{{\cal{X}}}$ and a particle ${\cal{Y}}$  at rapidity $y_{{\cal{Y}}}$ in an ultraperipheral $PbPb$ collision, with ${\cal{X}} \neq {\cal{Y}}$,  is given by \cite{klein,Klusek-Gawenda:2013dka,DSM}
\begin{eqnarray}
\frac{d^2\sigma \left[Pb Pb  \rightarrow   Pb \,\, {\cal{X}} \, {\cal{Y}} \,\, Pb \right]}{dy_{{\cal{X}}} dy_{{\cal{Y}}}} =  \int_{b_{min}}
\frac{d\sigma \,\left[Pb Pb \rightarrow   Pb \otimes {\cal{X}} \otimes Pb \right]}{d^2{\mathbf b} dy_{{\cal{X}}}}
\times 
\frac{d\sigma \,\left[Pb Pb \rightarrow   Pb \otimes  {\cal{Y}} \otimes Pb \right]}{d^2{\mathbf b} dy_{{\cal{Y}}}}
\,\, d^2{\mathbf b} \,\,,
\label{Eq:double}
\end{eqnarray} 
where ${\mathbf b}$ is the impact parameter of the collision. 
In order to consider only ultraperipheral collisions, we will assume  $b_{min} = 2 R_{Pb}$,  which is equivalent to treat 
the nuclei as hard spheres and that  excludes the overlap between the colliding hadrons. One has verified that similar results are obtained assuming  $b_{min} = 0$ and including the survival factor $P_{NH}(b)$ that describes the probability of no additional hadronic interaction between the nuclei, which is usually estimated using the Glauber formalism \cite{klein,Baltz:2009jk}. The symbol
$\otimes$ represents the presence of a rapidity gap in the final state.
In what follows, we will assume that ${\cal{X}}$ and ${\cal{Y}}$ will be produced by $\gamma \pom$ and $\gamma \gamma$ interactions, respectively. Considering  ${\cal{X}}$ as being a vector meson ($ J/\Psi$ or $\phi$), one has that \cite{upc}
\begin{eqnarray}
\frac{d\sigma \,\left[Pb Pb \rightarrow   Pb \otimes  {\cal{X}} \otimes Pb\right]}{d^2{\mathbf b} dy_{\cal{X}}} = \omega N_{Pb}(\omega,{\mathbf b})\,\sigma_{\gamma Pb \rightarrow {\cal{X}} \otimes Pb}\left(\omega \right)\,\,,
\label{dsigdy}
\end{eqnarray}
where the rapidity $y_{\cal{X}}$ of the vector meson in the final state is determined by the photon energy $\omega$ in the collider frame and by mass $M_{{\cal{X}}}$ of the vector meson [$y_{\cal{X}}\propto \ln \, ( \omega/M_{{\cal{X}}})$]. Moreover, $N_{Pb}(\omega,{\mathbf b})$ is  the equivalent photon spectrum associated with the ion, which will be discussed in more detail below. 
As in our previous study \cite{DSM}, we will estimate $\sigma_{\gamma Pb \rightarrow {\cal{X}} \otimes Pb}$ using the color dipole formalism, assuming the Gaus-LC model for the overlap function and the Glauber - Gribov model for the non-forward scattering  amplitude (See Refs. \cite{DSM,Azevedo:2022ozu} for details). It is important to emphasize that the associated results are able to describe the current data for the photoproduction of vector mesons in UPHICs (See, e.g. Ref. \cite{run2}).
 On the other hand, if ${\cal{Y}}$ is a system that can be produced by $\gamma \gamma$ interactions, one has that the associated differential cross section will be given by \cite{upc}
\begin{eqnarray}
\frac{d\sigma \,\left[Pb Pb \rightarrow   Pb \otimes  {\cal{Y}} \otimes Pb \right]}{d^2b dy_{{\cal{Y}}}}
&=& \int \mbox{d}^{2} {\mathbf B}\,
\mbox{d}W 
 \frac{W}{2} \, \hat{\sigma}\left(\gamma \gamma \rightarrow {\cal{Y}} ; 
W \right )  N\left(\omega_{1},{\mathbf b_{1}}  \right )
 N\left(\omega_{2},{\mathbf b_{2}}  \right )   
  \,\,\, , 
\label{cross-sec-2}
\end{eqnarray}
with ${\mathbf b} = ({\mathbf b_{1}} - {\mathbf b_{2}})$, ${\mathbf B} = ({\mathbf b_{1}} + {\mathbf b_{2}})/2$. Moreover,     $W = \sqrt{4 \omega_1 \omega_2}$ is the  $\gamma \gamma$ center - of - mass energy, the photon energies $\omega_i$ can be expressed in terms of $W$ and $y_{\cal{Y}}$ as follows: 
\begin{eqnarray}
\omega_1 = \frac{W}{2} e^{y_{\cal{Y}}} \,\,\,\,\mbox{and}\,\,\,\,\omega_2 = \frac{W}{2} e^{-y_{\cal{Y}}} \,\,\,,
\label{ome}
\end{eqnarray}
and $\hat{\sigma}$ is the cross section for the  $\gamma \gamma \rightarrow {\cal{Y}}$ subprocess.
If ${\cal{Y}}$ is a narrow resonance,  the Low formula \cite{Low:1960wv} can be used to express this cross section in terms of the two-photon decay width $\Gamma_{{\cal{Y}} \rightarrow \gamma \gamma}$ as  follows
\begin{eqnarray}
 \hat{\sigma}_{\gamma \gamma \rightarrow {\cal{Y}}}(\omega_{1},\omega_{2}) = 
8\pi^{2} (2J+1) \frac{\Gamma_{{\cal{Y}} \rightarrow \gamma \gamma}}{M_{\cal{Y}}} 
\delta(4\omega_{1}\omega_{2} - M_{{\cal{Y}}}^{2}) \, ,
\label{Low_cs}
\end{eqnarray}
where $M_{{\cal{Y}}}$ and $J$ are, respectively, the mass and spin of the  produced ${\cal{Y}}$ state. For ${\cal{Y}} = \eta_c$ we will assume the values of $\Gamma_{\eta_c \rightarrow \gamma \gamma}$, $J$ and $m_{\eta_c}$ as given in the PDG \cite{pdg}. In contrast,  for ${\cal{Y}}$ being a true muonium (TM) state, we will follows  Ref. \cite{Azevedo:2019hqp} and assume $M_{TM} = 2 m_{\mu}$ ($m_{\mu}$ is the muon mass), $J = 0$ and   that $\Gamma  = \alpha_{em}^5 m_{\mu}/2$, where  $\alpha_{em}$ is the electromagnetic coupling constant. As a consequence, one has that
\begin{eqnarray}
 \hat{\sigma}_{\gamma \gamma \rightarrow TM}(\omega_{1},\omega_{2}) = 2 \pi^2 \alpha_{em}^5 \delta(4\omega_{1}\omega_{2} - M_{TM}^{2})\,\,.
\end{eqnarray}
On the other hand, if ${\cal{Y}}$ is a dilepton system,  $\hat{\sigma}$ is the elementary cross section to produce a pair of
leptons with mass $m_l$ through the $\gamma \gamma \rightarrow l^+ l^-$ subprocess, which can be calculated using the Breit
- Wheller formula.  We will consider the dimuon production in what follows, which is satisfactorily described by the formalism  used in this letter, as demonstrated in Ref. \cite{Azevedo:2019fyz}.  
Finally, in order to estimate the differential cross sections we must specify the model assumed to calculate the 
the equivalent photon spectrum, $N(\omega_i, {\mathbf b}_i)$, which determines the number of photons with energy $\omega_i$ at a transverse distance ${\mathbf b}_i$  from the center of nucleus, defined in the plane transverse to the trajectory. Such  spectrum can be 
expressed in terms of the charge form factor $F(q)$ as follows \cite{upc}
\begin{eqnarray}
 N(\omega_i,b_i) = \frac{Z^{2}\alpha_{em}}{\pi^2}\frac{1}{b_i^{2} v^{2}\omega_i}
\cdot \left[
\int u^{2} J_{1}(u) 
F\left(
 \sqrt{\frac{\left( \frac{b_i\omega_i}{\gamma_L}\right)^{2} + u^{2}}{b_i^{2}}}
 \right )
\frac{1}{\left(\frac{b_i\omega_i}{\gamma_L}\right)^{2} + u^{2}} \mbox{d}u
\right]^{2} \,\,,
\label{fluxo}
\end{eqnarray}
where  $\gamma_L$ is the Lorentz factor and $v$ is the nucleus velocity.
In our analysis, we will consider the realistic form factor, which corresponds to the Wood - Saxon distribution and is the Fourier transform of the charge density of the nucleus, constrained by the experimental data. It can be analytically expressed by
\begin{eqnarray}
 F(q^{2}) = 
 \frac{4\pi\rho_{0}}{Aq^{3}} 
 \left[ 
 \sin(qR) - qR \cos(qR) 
 \right]
 \left[
 \frac{1}{1 + q^{2} a^{2}}
 \right]
\end{eqnarray}
with $a = 0.549$ fm and $R_{A} = 6.63$ fm \cite{DeJager:1974liz,Bertulani:2001zk}.

\begin{figure}[t]
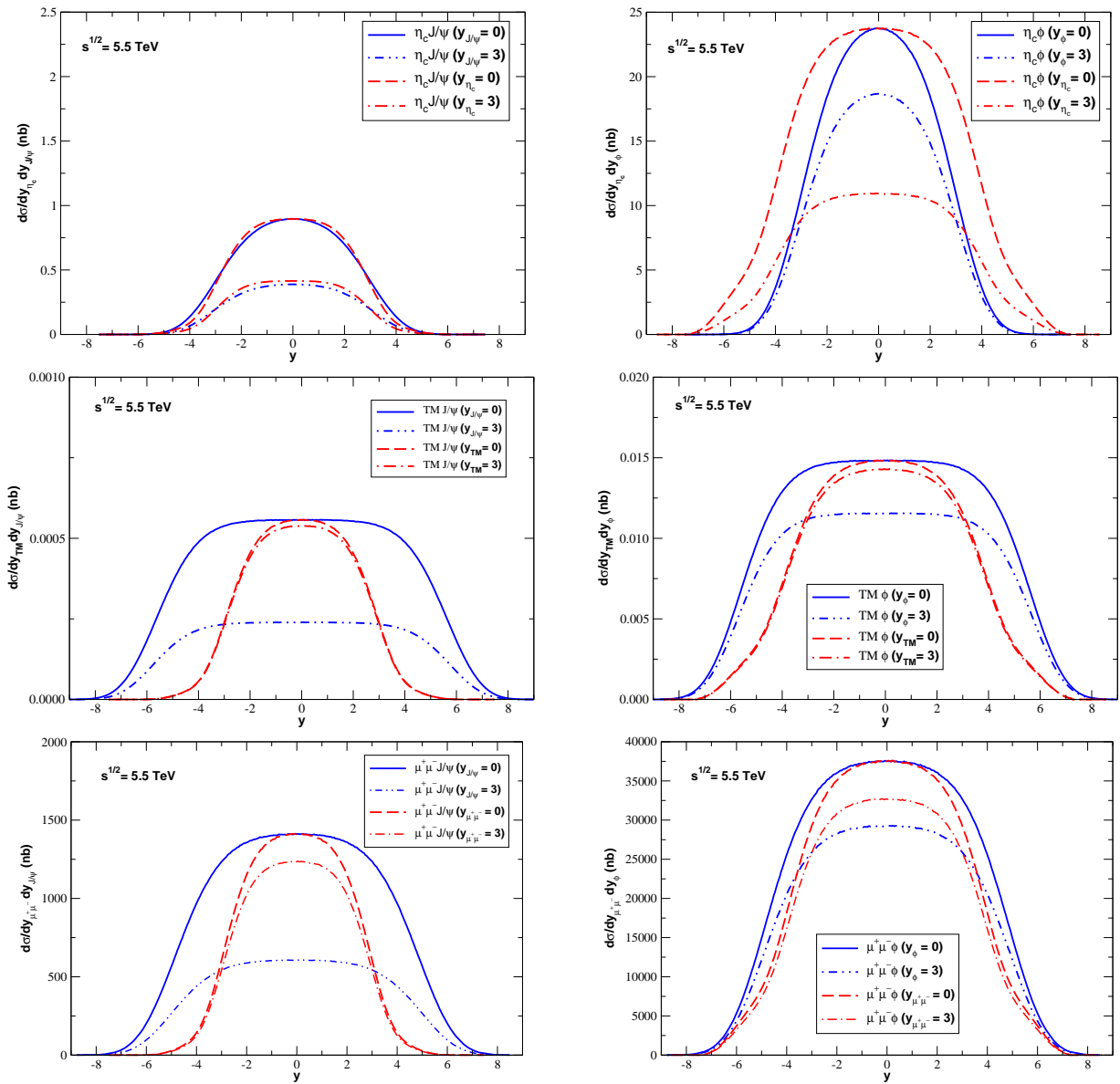

\begin{tabular}{ccc}
\includegraphics[scale=0.31]{rapidity_etacjpsi.eps} & \,\,\,\,\,\,\,\,\,\,\, &
\includegraphics[scale=0.31]{rapidity_etacphi.eps} \\
\includegraphics[scale=0.31]{rapidity_muoniumjpsi2.eps} & \,\,\,\,\,\,\,\,\,\,\, &
\includegraphics[scale=0.31]{rapidity_muoniumphi2.eps} \\
\includegraphics[scale=0.31]{rapidity_dimuonsjpsi.eps} & \,\,\,\,\,\,\,\,\,\,\, &
\includegraphics[scale=0.31]{rapidity_dimuonsphi.eps} 

\end{tabular}
\caption{Rapidity distributions for the double particle production via DSM in ultraperipheral $PbPb$ collisions at the LHC energy considering different combinations of final states. Results are derived assuming that the rapidity of one of the final states is constant and the dependence on the rapidity of the other particle is analyzed.}
\label{Fig:rapidity}
\end{figure}

Let's start our analysis presenting in Fig. \ref{Fig:rapidity} the predictions for the rapidity distributions associated with the double particle production via DSM in ultraperipheral $PbPb$ collisions at the LHC energy ($\sqrt{s} = 5.5$ TeV).  We consider  different combinations of final states and analyze the dependence on the rapidity of one of the particles in the final state, with the rapidity for the other particle being assumed constant. In particular, we assume two different values for the fixed rapidity, which are typical for central and forward detectors.  In the left (right) panels we present the predictions derived assuming that one of the particles is a $J/\Psi$ ($\phi$) meson.
One has that predictions involving a $J/\Psi$ meson have a smaller normalization and are narrower in rapidity than those with a $\phi$ meson. In general, the normalization is a factor $\gtrsim 20$ smaller. Moreover, for a given final state, the normalization of distribution decreases when a larger value of the fixed rapidity is considered. 
Similar conclusions are obtained for the FCC energy, where one has verified that the predictions are characterized by  larger normalizations and wider rapidity distributions. These results can be provided upon request.

The total cross sections for the double particle production via DSM in ultraperipheral $PbPb$ collisions are presented in Table \ref{Tab:secao-de-choque} considering the LHC and FCC energies as well as different rapidity ranges for the produced particles. We present predictions for the full rapidity ranges of LHC and FCC, as well as assuming that both particles are detected by a central ($-2.5  <y_{{\cal{X}},{\cal{Y}}}<2.5$) or a forward ($2.0  <y_{{\cal{X}},{\cal{Y}}}< 4.5$) detector, as e.g. the ALICE and LHCb detectors, respectively. One has that the predictions increase with the center - of - mass energy and  are lower when we impose that the particles are produced in rapidity range covered by a forward  detector. For the associated production of a vector meson with a $\eta_c$ meson, one has the prediction for the $\phi + \eta_c$ production are almost two orders of magnitude larger than those obtained for the $J/\Psi + \eta_c$ case. In comparison to the results presented in Ref. 
\cite{Azevedo:2022ozu}, where the $\phi + J/\Psi$ production cross section was estimated, our predictions for $\phi + \eta_c$ production are smaller by a factor $\approx 10^2$. Assuming that the  integrated luminosity expected for future runs of heavy ion collisions at the LHC and FCC will be  ${\cal{L}} \approx 7$ nb$^{-1}$, the  number of $\phi + \eta_c$ events  will be $\ge 10^3 \, (10^2)$ considering the central (forward) rapidity range. On the other hand, for the  
$\phi + TM$ production, our predictions are three order of magnitude smaller than those for the
$\phi + \eta_c$ case, which implies that the measurement of this final state in future runs will be a hard task. 
In contrast, the $\phi + \mu^+ \mu^-$ production cross section is four (six) orders of magnitude larger than the predictions for the $\phi + \eta_c$ ($\phi + TM$) case, implying that the associated number of events will be $\ge 10^7 \, (10^6)$ for the central (forward) selection. Although these numbers are reduced by two orders of magnitude if we consider the $J/\Psi + \mu^+ \mu^-$ production, a future experimental of this final state also is, in principle, feasible.

\begin{table}[t]
\begin{center}

\begin{tabular}{||c|c|c|c|c||}\hline \hline
 {\bf Final state}        & {\bf Rapidity range}        & \bf{LHC} ($\sqrt{s} = 5.02$ TeV) & \bf{LHC} ($\sqrt{s} = 5.5$ TeV)  & \bf{FCC} ($\sqrt{s} = 39$ TeV)   \\ \hline \hline
{\bf $J/\Psi + \eta_c$} & Full rapidity range            & $27.05$               & $29.72$               & $149.79$ \\ \hline
\,     & $-2.5<y_{J/\Psi,\eta_c}<2.5$    & $17.14$               & $18.19$               & $40.47$ \\ \hline
\,     & $\;\;2.0<y_{J/\Psi,\eta_c}<4.5$ & $0.72$                & $0.86$                & $8.26$ \\ \hline \hline 
{\bf $\phi + \eta_c$} & Full rapidity range          & $10.39 \times 10^2$   & $11.07 \times 10^2$  & $32.03 \times 10^2$ \\ \hline
\,     & $-2.5<y_{\phi, \eta_c}<2.5$    & $4.95 \times 10^2$    & $5.09 \times 10^2$   & $7.18 \times 10^2$ \\ \hline
\,     & $\;\;2.0<y_{\phi, \eta_c}<4.5$ & $0.37 \times 10^2$    & $0.41 \times 10^2$   & $1.47 \times 10^2$ \\ \hline \hline
{\bf $J/\Psi + TM$} & Full rapidity range            & $3.19 \times 10^{-2}$ & $3.45 \times 10^{-2}$ & $1.40 \times 10^{-1}$ \\ \hline
\,     & $-2.5<y_{J/\Psi, TM}<2.5$    & $1.20 \times 10^{-2}$ & $1.26 \times 10^{-2}$ & $2.45 \times 10^{-2}$ \\ \hline
\,     & $\;\;2.0<y_{J/\Psi, TM}<4.5$ & $1.07 \times 10^{-3}$ & $1.20 \times 10^{-3}$ & $5.70 \times 10^{-3}$ \\ \hline \hline
{\bf $\phi + TM$} & Full rapidity range            & $1.23$                & $1.29$          & $2.99$ \\ \hline
\,     & $-2.5<y_{\phi, TM}<2.5$    & $3.52 \times 10^{-1}$ & $3.56 \times 10^{-1}$ & $4.35 \times 10^{-1}$ \\ \hline
\,     & $\;\;2.0<y_{\phi, TM}<4.5$ & $5.66 \times 10^{-2}$ & $5.93 \times 10^{-2}$ & $1.03 \times 10^{-1}$ \\  \hline \hline 
{\bf $J/\Psi + \mu^{+}\mu^{-}$}     & Full rapidity range                        & $6.86 \times 10^{4}$ & $7.43 \times 10^{4}$ & $3.15 \times 10^{5}$ \\ \hline
\,     &  $-2.5 < y_{J/\Psi, \mu^{+}\mu^{-}} < 2.5$    & $2.99 \times 10^{4}$ & $3.14 \times 10^{4}$ & $6.24 \times 10^{4}$ \\ \hline
\,     & $\;\;2.0 < y_{J/\Psi, \mu^{+}\mu^{-}} < 4.5$ & $2.37 \times 10^{3}$ & $2.68 \times 10^{3}$ & $1.43 \times 10^{4}$ \\ \hline \hline
{\bf $\phi + \mu^{+}\mu^{-}$} & Full rapidity range                      & $2.65 \times 10^{6}$ & $2.78 \times 10^{6}$ & $6.75 \times 10^{6}$ \\ \hline
\,     & $-2.5 < y_{\phi, \mu^{+}\mu^{-}} < 2.5$    & $8.75 \times 10^{5}$ & $8.89 \times 10^{5}$ & $1.10 \times 10^{6}$ \\ \hline
\,     & $\;\;2.0 < y_{\phi, \mu^{+}\mu^{-}} < 4.5$ & $1.24 \times 10^{5}$ & $1.32 \times 10^{5}$ & $2.56 \times 10^{5}$ \\ \hline \hline
\end{tabular} 
\caption{Total cross sections (in nb) for the double particle production via DSM in ultraperipheral $PbPb$ collisions at the LHC and FCC energies considering different combinations of final states and distinct rapidity ranges for the produced particles.}
\label{Tab:secao-de-choque}
\end{center}
\end{table}

%\begin{figure}[t]
%\begin{tabular}{ccc}
%\includegraphics[scale=0.31]{energia_jpsieta.eps} & \,\,\,\,\,\,\,\,\,\,\, &
%\includegraphics[scale=0.31]{energia_phieta.eps} \\
%\includegraphics[scale=0.31]{energia_jpsimuonium2.eps} & \,\,\,\,\,\,\,\,\,\,\, &
%\includegraphics[scale=0.31]{energia_phimuonium2.eps} \\
%\includegraphics[scale=0.31]{energia_jpsidimuons.eps} & \,\,\,\,\,\,\,\,\,\,\, &
%\includegraphics[scale=0.31]{energia_phidimuons.eps} \\
%\end{tabular}
%\includegraphics[scale=0.42]{minv_ea_central_Pbp.pdf}
%\includegraphics[scale=0.42]{minv_ea_frontal_Pbp.pdf}
%\caption{Rapidity distributions for the $\phi J/\Psi$ production in ultraperipheral $PbPb$ collisions at the LHC and FCC energies considering the rapidity ranges covered by the ALICE (upper panels) and LHCb (lower panels) detectors. In the left (right) panels, the rapidity of the $J/\Psi$ ($\phi$) meson is assumed constant and the dependence on the rapidity of the $\phi$ ($J/\Psi$) meson is analyzed.}
%\label{Fig:energy}
%\end{figure} 

As a summary, one has that in recent years, several studies have demonstrated that the cross sections for the double vector meson production via the double scattering mechanism in ultraperipheral heavy ion collisions are huge and can be used to improve our understanding of the QCD dynamics at high energies. 
Such results have motivated the analysis performed in this letter, where we have estimated, for the first time, the associated production of a vector meson with a (pseudo)scalar state or a dimuon system in ultraperipheral $PbPb$ collisions for the LHC and FCC energies and presented predictions for the total cross sections and rapidity distributions considering the phase space covered by the ALICE and LHCb Collaborations. 
For the associated production of a vector meson with $\eta_c$ meson or a dimuon system, we predict large values for the total cross sections and for the number of events in future runs of the LHC and FCC, which indicate that the study of these final states are, in principle, feasible. In contrast, a future study of the associated production of a vector meson with a true muonium will be almost impossible considering the expected integrated luminosities. 
 Considering the results presented in this letter and those in Refs. \cite{DSM,Azevedo:2022ozu}, we strongly motivate a future investigation of the double particle production in ultraperipheral heavy ion collisions in order to probe the double scattering mechanism and improve our understanding of the QCD dynamics at high energies and meson structure.

\begin{acknowledgments}
V.P.G. thanks Murilo S. Rangel (UFRJ/Brazil and LHCb Collaboration) for useful discussions. This work was partially supported by CNPq, CAPES, FAPERGS and  INCT-FNA (Process No. 464898/2014-5).

\end{acknowledgments}

\hspace{1.0cm}

\end{document}